\documentclass[12pt]{article}
\begin{document}
\newfont{\elevenmib}{cmmib10 scaled\magstep1}%
\renewcommand{\theequation}{\arabic{section}.\arabic{equation}}
\newcommand{\tabtopsp}[1]{\vbox{\vbox to#1{}\vbox to12pt{}}}

\newcommand{\preprint}{
            \begin{flushleft}
   \elevenmib Yukawa\, Institute\, Kyoto\\
            \end{flushleft}\vspace{-1.3cm}
            \begin{flushright}\normalsize  \sf
            YITP-99-47\\
            DAMTP-1999-135\\
   {\tt hep-th/9910033} \\ October 1999
            \end{flushright}}
\newcommand{\Title}[1]{{\baselineskip=26pt \begin{center}
            \Large   \bf #1 \\ \ \\ \end{center}}}
\newcommand{\Author}{\begin{center}\large \bf
            A.\,J.\, Bordner$^a$%
           \footnote{Current address:
    Department of Bioengineering,
    University of California, San Diego\\
    \phantom{Current address}\qquad\ \
    9500 Gilman Drive,
    San Diego, CA.  92093-0412, USA\\
\hspace*{0.7cm}%
e-mail: abordner@be-research.ucsd.edu},
          N.\,S.\, Manton$^b$%
\footnote{%
\hspace*{0.08cm}%
e-mail: N.S.Manton@damtp.cam.ac.uk}
     and R.\, Sasaki$^a$%
\footnote{%
\hspace*{0.08cm}%
e-mail: ryu@yukawa.kyoto-u.ac.jp} \end{center}}
\newcommand{\Address}{\begin{center} \it
            $^a$ Yukawa Institute for Theoretical Physics, Kyoto
            University,\\ Kyoto 606-8502, Japan \\
     $^b$ Department of Applied Mathematics and Theoretical Physics,
     University of
            Cambridge,\\ Silver Street, Cambridge CB3 9EW, United
Kingdom
      \end{center}}
\newcommand{\Accepted}[1]{\begin{center}{\large \sf #1}\\
            \vspace{1mm}{\small \sf Accepted for Publication}
            \end{center}}
\baselineskip=20pt

\preprint
\thispagestyle{empty}
\bigskip
\bigskip

\Title{Calogero-Moser Models V: \\ Supersymmetry
and Quantum Lax Pair}
\Author

\Address
\vspace{1.5cm}

\begin{abstract}
It is shown that the
Calogero-Moser models based on all root systems  of the finite
reflection  groups (both the crystallographic and
non-crystallographic cases)
with the rational (with/without a harmonic confining potential),
trigonometric
and hyperbolic potentials can be simply supersymmetrised in terms of
superpotentials.
There is a universal formula for the supersymmetric ground state
wavefunction. Since the bosonic part of each supersymmetric model
is the usual quantum Calogero-Moser model,
this gives a universal formula for its ground state wavefunction and
energy, which is determined purely algebraically.
Quantum Lax pair operators and conserved quantities for all the above
Calogero-Moser models are established.
\end{abstract}
\bigskip
\bigskip
\bigskip

\section{Introduction}
\label{intro}
\setcounter{equation}{0}

The supersymmetric generalisation of quantum Calogero-Moser models in terms
of
superpotentials is presented.
It applies to all of the Calogero-Moser models based on the crystallographic
and non-crystallographic root systems and with the degenerate potentials,
{\em i.e.} the rational, hyperbolic and trigonometric potentials.
The supersymmetric ground state is easy to obtain and has zero energy,
and we are able to deduce a universal formula for the ground state
wavefunction and ground
state energy of the non-supersymmetric models. Our calculations
involve the consideration of all the two-dimensional sub-root systems
lying in the original one.
Historically, the integrability of Calogero-Moser models \cite{CalMo,OP2}
was
first discovered in the quantum mechanical models.
As we will show in this paper, the quantum and classical integrability
\cite{bcs1,bcs2,DHoker_Phong} are very closely related.
In this paper, the generic case with the elliptic potentials will not
be discussed. Supersymmetrisation and quantisation of Calogero-Moser models
with elliptic potentials remains a great challenge.

For general background and the motivations for this
paper, and for the physical applications of the
Calogero-Moser models with various potentials to
lower-dimensional physics, ranging from solid state to
particle physics and supersymmetric gauge theories, we refer to our
previous papers \cite{bcs1,bcs2} and references therein.

This paper is organised as follows. In section 2 we summarise the classical
Calogero-Moser models to set the stage and introduce appropriate notation.
One special property (\ref{Mtilsum}) of the Lax pair for the models with
degenerate potentials is pointed out. This property will be essential
in constructing the quantum Lax pair operators in section 4.
In section 3 the supersymmetrisation of quantum Calogero-Moser models
with degenerate potentials is presented, and
we derive the formulas for the ground state
wavefunction (\ref{grsol}) and the ground state energy (\ref{e0fom}),
(\ref{ratgren}) of
the non-supersymmetric models. In section 4 we derive quantum Lax pair
equations (\ref{qlaxL}) and (\ref{qlaxLcal}) for the non-supersymmetric
models and deduce the quantum conserved quantities (\ref{qconv}) and
(\ref{qconvcal}). Section 5 is for comments and discussion.

\section{Calogero-Moser Models}
\label{cal-mo}
\setcounter{equation}{0}
In this section we briefly introduce the {\em classical} Calogero-Moser
models along with
appropriate notation and background for the main body of this paper.
We consider only the degenerate potentials, that is the rational
(with/without
harmonic force), hyperbolic and trigonometric potentials.
In these cases the universal Lax pair operator \cite{bcs2} {\em without
spectral parameter} is drastically simplified and some new features not
shared
by the most general Lax pair arise. These will become important for the
{\em quantum} Lax pairs and conserved quantities to be discussed
in section \ref{qlax}.

\subsection{Model}
A  (generalised) Calogero-Moser model is a
Hamiltonian system associated with a root system $\Delta$
of rank \(r\),
which is a set of
vectors in $\mathbf{R}^{r}$ with its standard inner product,
invariant under reflections
in the hyperplane perpendicular to each
vector in $\Delta$.  In other words,
\begin{equation}
   s_{\alpha}(\beta)\in\Delta,\quad\forall \alpha,\beta\in\Delta,
\end{equation}
where
\begin{equation}
   s_{\alpha}(\beta)=\beta-2(\alpha\cdot\beta/|\alpha|^{2})\alpha.
\end{equation}
Dual roots are defined by $\alpha^{\vee}=2\alpha/|\alpha|^{2}$, in
terms of which
\begin{equation}
   \label{Root_reflection}
   s_{\alpha}(\beta)=\beta-(\alpha^{\vee}\!\!\cdot\beta)\alpha.
\end{equation}
The set of reflections $\{s_{\alpha},\,\alpha\in\Delta\}$ generates a
group, known as a Coxeter group, or finite reflection group.
The orbit of $\beta\in\Delta$ is the set of root vectors
resulting from the action of the Coxeter group on it.
The set of positive roots $\Delta_{+}$ may be defined in terms of a
vector $U\in\mathbf{R}^{r}$, with
$\alpha\cdot U \neq 0,\,\forall\alpha\in\Delta$, as
those roots $\alpha\in\Delta$ such that $\alpha\cdot U>0$.  Given
$\Delta_{+}$, there is a unique
set of $r$ simple roots $\Pi = \{\alpha_{j},\,j=1,\ldots, r\}$
defined such that they span
the root space and the coefficients $\{a_{j}\}$ in
$\beta=\sum_{j=1}^{r}a_{j}\alpha_{j}$ for $\beta\in\Delta_{+}$
are all non-negative. The highest root  $\alpha_h$, for which
$\sum_{j=1}^{r}a_{j}$ is maximal, is then also determined
uniquely.
The subset of reflections $\{s_{\alpha},\,\alpha\in\Pi\}$ in fact generates
the
Coxeter group.  The products of
$s_{\alpha}$, with $\alpha\in \Pi$, are subject solely to the relations
\begin{equation}
   \label{Coxeter_relations}
   (s_{\alpha}s_{\beta})^{m(\alpha,\beta)}=1,\qquad \alpha,\beta\in \Pi.
\end{equation}
The interpretation is that $s_{\alpha}s_{\beta}$ is a rotation in some
plane by $2\pi/{m(\alpha,\beta)}$.
The set of positive integers $m(\alpha,\beta)$
(with $m(\alpha,\alpha)=1,\,\forall \alpha\in \Pi$)
uniquely specify the Coxeter group.

The root systems for finite reflection groups may be divided into two
types: crystallographic and non-crystallographic.
Crystallographic root systems satisfy the additional condition
\begin{equation}
   \alpha^{\vee}\!\!\cdot\beta\in\mathbf{Z},\quad \forall
   \alpha,\beta\in\Delta,
\end{equation}
which implies that the $\mathbf{Z}$-span of $\Pi$ is a lattice in
$\mathbf{R}^{r}$ and  contains all roots in $\Delta$.
These root systems are associated with simple Lie
algebras: \{$A_{r},\,r\ge 1\}$, $\{B_{r},\,r\ge 2\}$, $\{C_{r},\,r\ge
2\}$,
$\{D_{r},\,r\ge 4\}$, $E_{6}$, $E_{7}$, $E_{8}$, $F_{4}$ and
$G_{2}$, and also $\{BC_{r},\,r\ge 2\}$ which combines the root
systems $B_{r}$ and $C_{r}$.  The Coxeter groups for these root
systems are called Weyl groups.  The remaining non-crystallographic root
systems \cite{Coxeter_groups} are $H_{3}$, $H_{4}$, whose Coxeter groups are
the symmetry
groups of the icosahedron and four-dimensional 600-cell, respectively,
and the dihedral group of order $2m$, $\{I_{2}(m),\,m\ge 4\}$.

The dynamical variables of the Calogero-Moser model are the coordinates
$\{q_{j}\}$ and their canonically conjugate momenta $\{p_{j}\}$, with
the Poisson brackets
\begin{equation}
   \{q_{j},p_{k}\}=\delta_{jk},\qquad \{q_{j},q_{k}\}=
   \{p_{j},p_{k}\}=0,\quad j,k=1,\ldots,r.
\end{equation}
These will be denoted by vectors in $\mathbf{R}^{r}$
\begin{equation}
   q=(q_{1},\ldots,q_{r}),\qquad p=(p_{1},\ldots,p_{r}).
\end{equation}
The Hamiltonian for the {\em classical} Calogero-Moser model is
\begin{equation}
   \label{cCMHamiltonian}
   \mathcal{H}_C = {1\over 2} p^{2} + {1\over2}\sum_{\rho\in\Delta_+}
   {g_{|\rho|}^{2} |\rho|^{2}}
   \,V(\rho\cdot q),
\end{equation}
in which the real {\em positive} coupling constants $g_{|\rho|}$
 are defined on orbits of the corresponding
Coxeter group, {\it i.e.} they are
identical for roots in the same orbit.
That is, for the simple Lie algebra cases
\(g_{|\rho|}=g\) for all roots in simply-laced models
and  \(g_{|\rho|}=g_L\)
for long roots and \(g_{|\rho|}=g_S\) for
short roots in non-simply laced models. For the $BC_r$ models there
 are three couplings, and in the $I_{2}(m)$ models,
 there is one coupling if $m$ is odd, and two if $m$ is even (see
 section \ref{susy}).
The $H_3$ and $H_4$ models have one coupling constant \(g_{|\rho|}=g\),
since these root systems are simply-laced.
(Exhibiting the factor $|\rho|^{2}$, rather than absorbing it
into the coupling constant, is a convenience.)
This then ensures that for any potential $V$,
the Hamiltonian is invariant under reflection of the phase space
variables in the hyperplane perpendicular to any root
\begin{equation}
   q\rightarrow s_{\alpha}(q), \quad p\rightarrow s_{\alpha}(p), \quad
   \forall\alpha\in\Delta
\end{equation}
with $s_{\alpha}$ defined by (\ref{Root_reflection}).

The Lax pair operators that  we will introduce will apply
for the following degenerate potentials:
\begin{equation}
   V(\alpha\cdot q)=
   \cases{{1/{(\alpha\cdot q)^2}},
   \cr\cr
   {a^2/{\sinh^2 a(\alpha\cdot q)}},
   \cr\cr {a^2/{\sin^2 a(\alpha\cdot q)}},
   \cr}
   \label{potfun}
\end{equation}
in which \(a\) is an arbitrary real positive constant, determining the
period of the trigonometric potentials (and the imaginary period in
the hyperbolic case, although this has less significance).
They imply integrability for all of the
Calogero-Moser models based on the crystallographic root systems.
Those models based on the non-crystallographic root systems,  the dihedral
group
\(I_2(m)\), \(H_3\), and \(H_4\), are integrable only for  the
rational potential.
The rational potential models are also integrable if a confining
harmonic
potential
\begin{equation}
   {1\over2}\omega^2q^2,\quad \omega>0
   \label{harmpot}
\end{equation}
is added to the Hamiltonian.

Some remarks are in order. For all of the root systems and for
any choice of potential (\ref{potfun}),
the  Calogero-Moser model has a hard repulsive potential \(\sim
{1/{(\alpha\cdot q)^2}}\) near the reflection hyperplane
\(H_{\alpha}=\{q\in\mathbf{R}^{r},\, \alpha\cdot q=0\}\).
The strength of the singularity is given by
the coupling constant \(g_{|\alpha|}^2\)
which is {\em independent} of the choice of the normalisation of the
roots.
This determines the form of the ground state wave function
in the quantum version of the theory, as we will
see in section \ref{susy}.
The repulsive potential is classically insurmountable.
Thus the motion is always
confined within one Weyl chamber.
This  feature allows us to constrain the configuration space to
the principal Weyl chamber
\begin{equation}
   PW=\{q\in{\bf R}^r|\ \alpha\cdot q>0,\quad \alpha\in\Pi\},
   \label{PW}
\end{equation}
without loss of generality.
In the case of the trigonometric potential, the configuration space is
further
limited due to the periodicity of the potential to
\begin{equation}
   PW_T=\{q\in{\bf R}^r|\ \alpha\cdot q>0,\quad \alpha\in\Pi,
   \quad \alpha_h\cdot q<\pi/a\},
   \label{PWT}
\end{equation}
where \(\alpha_h\) is the highest root.

\subsection{Lax Pair}
Here we recapitulate the essence of the universal Lax pair operators
for the Calogero-Moser models with degenerate potentials and without
spectral parameter.
For details and a full exposition, see \cite{bcs2}.
The Lax operators are
\begin{eqnarray}
   L &=& p\cdot\hat{H}+X,\qquad X=i\sum_{\rho\in\Delta_{+}}g_{|\rho|}
   \,\,(\rho\cdot\hat{H})\,x(\rho\cdot q)\,\hat{s}_{\rho},
   \label{LaxOpDef}\\
   \widetilde{M} &=&
   {i\over2}\sum_{\rho\in\Delta_{+}}g_{|\rho|}|\rho|^2\,y
   (\rho\cdot q)\,\hat{s}_{\rho},
   \label{Mtildef}
\end{eqnarray}
in which $\{\hat{s}_{\alpha},\,\alpha\in\Delta\}$ are the
reflection operators  of the root system.
They act on a set of $\mathbf{R}^{r}$ vectors
\({\cal R}=\{\mu^{(k)}\in\mathbf{R}^{r},\ k=1,\ldots, d\}\), permuting them
under the action of the reflection group.
The vectors in $\cal R$ form a basis for
the representation space $\bf V$ of dimension $d$.
The simplest and the most natural representation spaces of the Lax
pair operators are provided by the set of all roots $\Delta$ for
the simply-laced root systems, and the set of short roots $\Delta_S$ or
the set of long roots $\Delta_L$ for non-simply laced root systems.
These give root type Lax pairs, \cite{bcs1}.
Another class of simple representations are the so-called minimal type
representations, for which ${\cal R}$ consists of
the weights belonging to a minimal representation, and
which give minimal type Lax pairs \cite{bcs1}.

The set of operators $\{\hat{H}_{j},\,j=1,\ldots, r\}$ are defined as
follows.
If  $\hat{H}_{j}$ acts on a vector $\mu^{(k)}\in{\cal R}$, the $j$-th
component is returned:
\[
\hat{H}_{j}\mu^{(k)}=\mu^{(k)}_j\mu^{(k)}.
\]
These, along with the reflection operators, form the following operator
algebra:
\begin{eqnarray}
   \label{OpAlgebra1}
   [\hat{H}_{j},\hat{H}_{k}]=0, \\
   \label{OpAlgebra2}
   [\hat{H}_{j},\hat{s}_{\alpha}] = \alpha_{j}
   (\alpha^{\vee}\!\!\cdot\hat{H})\hat{s}_{\alpha},
   \\
   \label{OpAlgebra3}
   \hat{s}_{\alpha}\hat{s}_{\beta}\hat{s}_{\alpha}
   =\hat{s}_{s_{\alpha}(\beta)},
   \\
   \label{OpAlgebra4}
   (\hat{s}_{\alpha}\hat{s}_{\beta})^{m(\alpha,\beta)}=1.
\end{eqnarray}
The first relation (\ref{OpAlgebra1}) implies that the
operators $\{\hat{H}_{j},\,j=1,\ldots,
r\}$ form an abelian subalgebra and relations (\ref{OpAlgebra3}) and
(\ref{OpAlgebra4}) are just
those for the finite reflection group associated with the root system
$\Delta$.  The set of integers $m(\alpha,\beta)$ are those appearing
in the Coxeter relations
(\ref{Coxeter_relations}) which characterise the reflection group.

The form of the function $x$ depends on the chosen potential, and
the function $y$ and another function $w$ to be used in section
\ref{susy} are defined by
\begin{equation}
   y(u)\equiv {d\over {du}} x(u),\quad
   {dw(u)\over {du}}/w(u)\equiv x(u).
   \label{hdef}
\end{equation}
They have definite parities:
\begin{equation}
   x(-u)=-x(u),\quad  y(-u)=y(u),\quad  w(-u)=-w(u),
   \label{xodd}
\end{equation}
so that $L$ and
$\widetilde{M}$ are independent of the choice of positive roots
$\Delta_{+}$.
This also implies that the sums in (\ref{LaxOpDef}), (\ref{Mtildef}) may
be extended to a sum over all roots if an additional factor of $1/2$ is
included in front of
the sums since the summands are
even under $\rho\rightarrow -\rho$.
The functions $x$ and $y$ are further related to each other,
and to the potential function $V$ occurring in the Hamiltonian via
\begin{equation}
   V(u)=-y(u)=x^2(u)+a^2\times\left\{
   \begin{array}{rl}
      0& \mbox{rational}\\
      -1& \mbox{hyperbolic}\\
      1& \mbox{trigonometric.}
   \end{array}\right.
   \label{Vxrel}
\end{equation}
Note that these relations are only valid for the degenerate potentials
(\ref{potfun}) and
in the Lax pair without spectral parameter.
The following Table 1 gives these functions for each potential:
\begin{center}
 \begin{tabular}{|l|c|c|c|}
 \hline
  & $w(u)$ & $x(u)$ & $y(u)$ \\
 \hline
 rational & $u$ & $1/u$ & -$1/u^2$ \\
 \hline
 hyperbolic & $\sinh au$ & $a\coth au$ & -$a^2/\sinh^2 au$ \\
 \hline
 trigonometric & $\sin au$ & $a\cot au$ & -$a^2/\sin^2 au$ \\
 \hline
 \end{tabular}\\
 \bigskip
 Table 1: Functions appearing in the Lax pair and superpotential.
\end{center}
The underlying idea of the Lax operator $L$, (\ref{LaxOpDef}), is quite
simple.
As seen from (\ref{L2Ham}), $L$ is a ``square root"
of the Hamiltonian.
Thus one part of $L$ contains $p$ which is not associated with
roots and another part contains \(x(\rho\cdot q)\), a ``square root"
of the potential
\(V(\rho\cdot q)\), which being associated with a root $\rho$
is therefore accompanied by the reflection operator $\hat{s}_{\rho}$.

It is straightforward to show that the Lax equation
\begin{equation}
   \label{LaxEquation}
   {d\over dt}{L}=[L,\widetilde{M}],
\end{equation}
which divides into two parts as
\begin{eqnarray}
   \label{LaxEqn1}{d\over dt}X &=& [p\cdot\hat{H},\widetilde{M}], \\
   \label{LaxEqn2}{d\over dt}(p\cdot\hat{H}) &=&
   [X,\widetilde{M}],
\end{eqnarray}
is equivalent to the canonical equations of motion:
\begin{eqnarray}
   \dot{q}_{j} &=& {\partial{\cal H}_C\over{\partial p_{j}}} =  p_{j},\\
   \label{EqnsOfMotion}
   \dot{p}_{j} &=& -{\partial{\cal H}_C\over{\partial q_{j}}}
   =-{\partial\over \partial q_{j}}\left[\sum_{\rho\in\Delta_+}
   {1\over2}{g_{|\rho|}^{2} |\rho|^{2}}V(\rho\cdot q)\right].
   \label{eqMot2}
\end{eqnarray}
For the details of the proof, see \cite{bcs2}.
It is amusing to note that the Lax equation is rather symmetric in
$X\leftrightarrow p\cdot\hat{H}$.
In section \ref{qlax} we will discuss the quantum version of these
equations.

It is well-known that conserved quantities are given in terms of
a representation ${\cal R}$ of the operator $L$ as
\begin{equation}
   \mbox{Tr}(L^n)\equiv\sum_{\mu\in{\cal R}}(L^n)_{\mu\mu},\quad
   n=1,2,\ldots,
\end{equation}
in which $\mu$'s are the basis vectors of the representation ${\cal
R}$.
In particular, the classical Hamiltonian (\ref{cCMHamiltonian})
is given by
\begin{equation}
   {\cal H}_C = {1\over {2 C_{\cal R}}}\mbox{Tr}(L^{2}) +
   const,
   \label{L2Ham}
\end{equation}
where the constant $C_{\cal R}$, which depends on the
 representation, is
defined by
\begin{equation}
   \mbox{Tr}(\hat{H}_{j}\hat{H}_{k})
   \equiv\sum_{\mu\in{\cal R}}(\hat{H}_{j}\hat{H}_{k})_{\mu\mu}
   =\sum_{\mu\in{\cal R}}\mu_{j}\mu_{k}
   =C_{\cal R}\,\delta_{jk}.
\end{equation}

Before closing this section let us remark on one special property of a
representation matrix of the Lax operator $\widetilde{M}$:
\begin{equation}
   \sum_{\mu\in{\cal R}}\widetilde{M}_{\mu\nu}=
   \sum_{\nu\in{\cal R}}\widetilde{M}_{\mu\nu}=
   -i{\cal D}\equiv{i\over2}\sum_{\rho\in\Delta_{+}}g_{|\rho|}|\rho|^2\,y
      (\rho\cdot q)
   =-{i\over2}\sum_{\rho\in\Delta_+}g_{|\rho|}|\rho|^2\,V
   (\rho\cdot q).
   \label{Mtilsum}
\end{equation}
The quantity ${\cal D}$ is independent of the representation ${\cal R}$.
Thus we can define a new Lax operator $M$ by
\begin{equation}
   M=\widetilde{M}+i{\cal D}\times I,\quad I:\ \mbox{Identity operator},
   \label{Mdefnew}
\end{equation}
which satisfies the relation
\begin{equation}
   \sum_{\mu\in{\cal R}}M_{\mu\nu}=
   \sum_{\nu\in{\cal R}}M_{\mu\nu}=0.
   \label{sumMzero}
\end{equation}
The new Lax pair $L$ and $M$ gives the same {\em classical}
equations of motion as above.
The above property (\ref{sumMzero}) has been known for the $A_r$ model Lax
pair in the vector representation \cite{ShaSu}.
We stress that this is a universal property shared by all of the Lax
matrices
without spectral parameter for the degenerate potentials in any
representation. The reason is that the
$\mu\nu$ matrix element of the $\widetilde{M}$ operator
(\ref{Mtildef}) reads
\begin{equation}
\widetilde{M}_{\mu\nu}=
   {i\over2}\sum_{\rho\in\Delta_{+}}g_{|\rho|}|\rho|^2\,y
   (\rho\cdot q)\,(\hat{s}_{\rho})_{\mu\nu},
\end{equation}
in which
\begin{equation}
   (\hat{s}_{\rho})_{\mu\nu}=\delta_{\mu,s_\rho(\nu)}=
   \delta_{\nu,s_\rho(\mu)}.
\end{equation}
Since $s_\rho(\nu)$ ($s_\rho(\mu)$) is always contained in the basis
of the representation precisely once,
\begin{equation}
   \sum_{\mu\in{\cal R}}(\hat{s}_{\rho})_{\mu\nu}=
   \sum_{\nu\in{\cal R}}(\hat{s}_{\rho})_{\mu\nu}=1
   \label{ssumform}
\end{equation}
and so (\ref{Mtilsum}) is obtained.
This also means that for $\mu\neq\nu$, $\widetilde{M}_{\mu\nu}$ is
either 0
or it consists of a single term (two terms in the \(BC_r\) case).
For suppose two positive roots
$\rho$ and $\sigma$ connect $\mu$ and $\nu$, then we obtain
\begin{eqnarray*}
   \mu&=&s_{\rho}(\nu)=\nu-(\rho^\vee\!\!\cdot\nu)\rho\\
   &=&s_{\sigma}(\nu)=\nu-(\sigma^\vee\!\!\cdot\nu)\sigma.
\end{eqnarray*}
This would imply $\rho\propto \sigma$ which then means $\rho=\sigma$
(\(\rho=2\sigma\) or \(\sigma=2\rho\) in the  \(BC_r\) case)
since both
are positive roots.
The diagonal element $\widetilde{M}_{\mu\mu}$ contains
contributions from all of the roots which are orthogonal to $\mu$.

\subsubsection{Rational potential with harmonic force}
Here we give
the Lax pair for the  rational potential model with harmonic force.
The Hamiltonian is
\begin{equation}
   {\cal H}_{C\omega}={1\over2}p^2+{1\over2}\omega^2q^2+
   {1\over2}\sum_{\rho\in\Delta_+}
      {g_{|\rho|}^{2}}{|\rho|^{2}\over{(\rho\cdot q)^2}}.
   \label{1stratharm}
\end{equation}
The canonical equations of motion are equivalent to the following
Lax equations for $L^{\pm}$:
\begin{equation}
   \dot{L}^{\pm}=[L^{\pm},\widetilde{M}]\pm i\omega L^{\pm},
   \label{omegaLM}
\end{equation}
in which (see section 4 of \cite{bcs2})
$\widetilde{M}$ is the same as before (\ref{Mtildef}), and
 $L^{\pm}$ and $Q$ are defined by
\begin{equation}
   L^{\pm}=L\pm i\omega Q, \quad Q=q\cdot\hat{H},
\end{equation}
with $L$, $\hat{H}$ as earlier.
If we define hermitian operators \({\cal L}_1\) and \({\cal L}_2\) by
\begin{equation}
   {\cal L}_1=L^+L^-,\quad {\cal L}_2=L^-L^+,
   \label{defcalL}
\end{equation}
they satisfy Lax-like equations
   \begin{equation}
   \dot{{\cal L}}_k=[{\cal L}_k,\widetilde{M}],\quad k=1,2.
\end{equation}
\noindent From these we can construct conserved quantities
\begin{equation}
   \mbox{Tr}({\cal L}_1^n)=\mbox{Tr}({\cal L}_2^n),\quad n=1,2,\ldots,
\end{equation}
as before. It is elementary to check that the first conserved quantities
give the Hamiltonian (\ref{1stratharm})
\begin{equation}
   \mbox{Tr}({\cal L}_1)= \mbox{Tr}({\cal L}_2)\propto {\cal
H}_{C\omega}+const.
\end{equation}
As in the other cases, the operator $\widetilde{M}$ can be replaced by
$M$, (\ref{Mdefnew}), without changing the {\em classical} equations of
motion.
This then completes the presentation of the Lax pairs for all of the
classical Calogero-Moser models with non-elliptic potentials.

\section{Supersymmetrisation}
\label{susy}
\setcounter{equation}{0}

\subsection{Superpotential and Hamiltonian}
In this section we show that all the Calogero-Moser models with
degenerate potentials summarised in the previous section can be simply
supersymmetrised in terms of superpotentials.
The result is a quantum system with bosonic and fermionic
variables. There are some pioneering works on supersymmetric
Calogero-Moser models with degenerate potentials, mainly those
based on $A_r$ and other classical root systems \cite{FM,ShaSu,Br}.
We shall not consider here the classical supersymmetric
Calogero-Moser models, which have dynamical variables
taking values in a Grassmann algebra, although these are interesting
too.

The bosonic variables have, as before, $2r$
degrees of freedom:
\begin{equation}
   q=(q_{1},\ldots,q_{r}),\qquad p=(p_{1},\ldots,p_{r}),
\end{equation}
with the canonical commutation relations
\begin{equation}
   [q_{j},p_{k}]=i\delta_{jk},\qquad [q_{j},q_{k}]=
   [p_{j},p_{k}]=0,\quad j,k=1,\ldots,r.
\end{equation}
The corresponding  ${\cal N}=2$ supersymmetric quantum
mechanical system requires additionally as many fermionic degrees of
freedom:
\begin{equation}
   \psi=(\psi_1,\ldots,\psi_r),\qquad \psi^*=(\psi^*_{1},\ldots,\psi^*_{r}),
\end{equation}
with the canonical anti-commutation relations
\begin{equation}
  \psi^{*}_j\psi_{k}+\psi_k\psi^{*}_j=\delta_{jk},\qquad
   \psi_{j}\psi_{k}+\psi_{k}\psi_{j}=
   \psi^{*}_j\psi^*_{k}+\psi^*_k\psi^{*}_j=0,\quad j,k=1,\ldots,r.
\end{equation}
The bosonic and fermionic variables commute with each other:
\begin{equation}
   [q_j, \psi_k]=[q_j, \psi^*_k]=[p_j, \psi_k]=[p_j, \psi^*_k]=0,\quad
   j,k=1,\ldots,r.
\end{equation}
We realize these variables as operators in the standard way, as acting
on wavefunctions which lie in the tensor product of the Hilbert space
of functions of $q$ and a $2^r$-dimensional fermionic Fock
space.  The momentum operator $p_j$ acts as
\[
   p_j=-i{\partial\over{\partial q_j}}, \quad j=1,\ldots,r.
\]
The fermionic variables $\psi$ and $\psi^*$ are respectively
annihilation and creation operators, which are hermitian conjugates of
each other. The bosonic variables $q$ will be restricted by the potential
in the same way as in the classical models to lie in the regions
(\ref{PW}) or (\ref{PWT}).

The dynamics of a supersymmetric quantum mechanical system
is determined by a {\em superpotential} \cite{EW}
$W(q)=W(q_1,\ldots,q_r)\in{\bf R}$.
The two {\em supercharges} ${\cal Q}$
and ${\cal Q}^*$ are defined by
\begin{equation}
   {\cal Q}=\sum_{j=1}^r\psi^{*}_j\left(p_j+i{\partial W\over{\partial
   q_j}}\right),
   \qquad
   {\cal Q}^*=\sum_{j=1}^r\psi_j\left(p_j-i{\partial W\over{\partial
   q_j}}\right),
\end{equation}
and the supersymmetric Hamiltonian is given by
\begin{equation}
   {\cal H}_{SUSY}={1\over2}\left({\cal Q}{\cal Q}^*+{\cal Q}^*{\cal
   Q}\right),
\end{equation}
which is obviously positive semi-definite.
They satisfy
\begin{equation}
   {\cal Q}^2={\cal Q}^{*2}=0,
   \qquad
   [{\cal H}_{SUSY},{\cal Q}]=[{\cal H}_{SUSY},{\cal Q}^*] =0.
\end{equation}

In terms of the superpotential, ${\cal H}_{SUSY}$ reads
\begin{eqnarray}
   {\cal H}_{SUSY}&=&{1\over 2}\sum_{j=1}^{r}\left(p_{j}^{2}+\left({\partial
W
   \over{\partial q_{j}}}\right)^{2}\right)-{1\over 2}\sum_{j,k=1}^{r}
   [\psi_{j}^{*},\psi_{k}]{\partial^{2}W\over{\partial q_{j}\partial
   q_{k}}},\\
   &=&{\cal H}_{B}+{\cal H}_{F},
   \label{HBHF}
\end{eqnarray}
in which the bosonic and fermionic parts are
\begin{eqnarray}
   {\cal H}_{B}&=&{1\over 2}\sum_{j=1}^{r}\left(p_{j}^{2}+\left({\partial W
   \over{\partial q_{j}}}\right)^{2}\right)
   +{1\over2}\sum_{j=1}^r{\partial^{2}W\over{\partial (q_{j})^2}},
   \label{HscB}\\
   {\cal H}_{F}&=&-\sum_{j,k=1}^{r}
   \psi_{j}^{*}\psi_{k}\,{\partial^{2}W\over{\partial q_{j}\partial
   q_{k}}}.\label{HSCF}
\end{eqnarray}
Note the ordering of the fermionic variables in ${\cal H}_{F}$, which
is responsible for the last term in ${\cal H}_{B}$.
The Calogero-Moser dynamics is specified by the following choice of the
superpotential
\begin{equation}
   W(q)=\sum_{\rho\in\Delta_+}g_{|\rho|}\ln|w(\rho\cdot q)|+
   (-{\omega\over2}q^2),
   \quad g_{|\rho|}>0,\quad \omega>0,
   \label{suppot}
\end{equation}
in which the function $w$ is defined by (\ref{hdef}) (see also Table
1), and the last term corresponds to the harmonic confining potential
in the rational potential
model, if present. It should be remarked that this superpotential is
{\em Coxeter-invariant}:
\begin{equation}
   \check{s}_{\rho}W=W,\qquad \forall\rho\in\Delta,
   \label{Coxinv}
\end{equation}
in which the new reflection operator $\check{s}_{\rho}$ acts on a
function $f$ of $q$ as follows
\begin{equation}
   (\check{s}_{\rho}f)(q)=f(s_{\rho}(q)).
   \label{defcheck}
\end{equation}

We show that the bosonic part, ${\cal H}_{B}$ of ${\cal H}_{SUSY}$,
(\ref{HscB}), can be written
as follows:
\begin{equation}
   {\cal H}_{B}={\cal H}_{C}+{\cal H}_{qc}-{\cal E}_0.
   \label{HBform}
\end{equation}
Here ${\cal H}_{C}$ is the classical Hamiltonian
(\ref{cCMHamiltonian}), interpreted as a quantum operator,
and ${\cal H}_{qc}$ is the ``quantum correction" term
derived from the last term of (\ref{HscB}):
\begin{eqnarray}
   {1\over2}\sum_{j=1}^r{\partial^{2}W\over{\partial
   (q_{j})^2}}
   &=&{1\over2}\sum_{\rho\in\Delta_+}g_{|\rho|}|\rho|^2\,y(\rho\cdot
   q)+(-{r\omega\over2} )\label{Hqcy1}\\
   &=&-{1\over2}\sum_{\rho\in\Delta_+}g_{|\rho|}|\rho|^2\,V(\rho\cdot
   q)+(-{r\omega \over2})\label{Hqcy2},\\
   {\cal
H}_{qc}&=&-{1\over2}\sum_{\rho\in\Delta_+}g_{|\rho|}|\rho|^2\,V(\rho\cdot
   q)= -{\cal D}.
   \label{Hqcform}
\end{eqnarray}
In deriving (\ref{Hqcy2}) from (\ref{Hqcy1}), the relation (\ref{Vxrel})
is used, and recall ${\cal D}$ is defined by (\ref{Mtilsum}).
The constant ${r\omega/2}$ becomes a part of the constant ${\cal E}_0$,
which  is the {\em ground state energy} of the bosonic Hamiltonian
${\cal H}_C+{\cal H}_{qc}$, since we shall see that both ${\cal H}_{SUSY}$
and
${\cal H}_{F}$ annihilate the ground state, so ${\cal H}_{B}$ must do so
too.

In order to show (\ref{HBform}) we need to evaluate
$\sum_j\left({\partial W /{\partial q_{j}}}\right)^{2}$.  Firstly we have
\[
   {\partial W
   \over{\partial q_{j}}}=\sum_{\rho\in\Delta_+}g_{|\rho|}
   {w^\prime(\rho\cdot q)\over{w(\rho\cdot q)}}\,\rho_j+
   (-{\omega}q_j)
   =\sum_{\rho\in\Delta_+}g_{|\rho|}x(\rho\cdot q)\,\rho_j+
   (-{\omega}q_j)
\]
so
\begin{eqnarray}
   \sum_j\left({\partial W
   \over{\partial q_{j}}}\right)^{2}&=&
   \sum_{\rho\in\Delta_+}
      {g_{|\rho|}^{2} |\rho|^{2}}
      \,x(\rho\cdot q)^2+(\omega^2q^2) \nonumber \\
   && +\sum_{\rho,\sigma\in\Delta_+\atop{\rho\neq\sigma}}\!\!
      {g_{|\rho|}g_{|\sigma|}(\rho\cdot\sigma})
      \,x(\rho\cdot q)\,x(\sigma\cdot q)+
   (-2\omega\!\!\sum_{\rho\in\Delta_+}\!\!
      g_{|\rho|}
      \,x(\rho\cdot q)\rho\cdot q).
   \label{HMmain}
\end{eqnarray}
The first line of (\ref{HMmain}) gives the potential terms of the
classical Hamiltonian (\ref{cCMHamiltonian}) up to a constant.
Secondly we show that the terms in the second line of (\ref{HMmain})
sum up to a constant depending on the root system and the choice of
potential.
The terms proportional to $\omega$ exist only for the rational potential
$x(u)=1/u$ and thus they give rise to a constant solely determined by the
root system:
\begin{equation}
   -2\omega\!\!\sum_{\rho\in\Delta_+}
   g_{|\rho|}
   \,x(\rho\cdot q)(\rho\cdot q)=-2\omega\!\!\sum_{\rho\in\Delta_+}
   g_{|\rho|}\cdot 1=
   -\omega\times\left\{\begin{array}{cl}
   gN& \mbox{simply-laced},\\
   g_SN_S+g_LN_L &\mbox{non-simply laced},
   \end{array}
   \right.
   \label{addharm}
\end{equation}
in which $N$ is the total number of roots in $\Delta$ and
$N_S$ ($N_L$) is the number of short (long) roots in $\Delta$.

The main task is to evaluate
  \begin{equation}
 \sum_{\rho,\sigma\in\Delta_+\atop{\rho\neq\sigma}}\!\!
   {g_{|\rho|}g_{|\sigma|}(\rho\cdot\sigma})
   \,x(\rho\cdot q)\,x(\sigma\cdot q),
\label{sums}
\end{equation}
where each distinct pair of roots gives two equal contributions.
This we do by decomposing it into
two-dimensional planes specified by two roots
$\rho$ and $\sigma$:
\begin{equation}
   f_{\Theta}(q)=\sum_{\rho\neq\sigma\in\Theta_{+},
   \,R_{\psi}=s_{\rho}s_{\sigma}}
   \!\!g_{|\rho|}g_{|\sigma|}(\rho\cdot\sigma)
   \,x(\rho\cdot q)\,x(\sigma\cdot q).
   \label{innerform}
\end{equation}
This quantity must be evaluated for a fixed sense of rotation
$R_{\psi}=s_{\rho}s_{\sigma}$ and all roots appearing in it are in the
two-dimensional sub-root system $\Theta=\{\kappa,\,\kappa\in(\Delta\cap
\mbox{span}(\rho,\sigma)\}$ with positive roots
$\Theta_{+}\equiv \Theta\cap \Delta_{+}$.
There is a reverse rotation $R_{-\psi}=s_{\sigma}s_{\rho}$ in the same plane
which gives the same contribution as $f_{\Theta}(q)$ as it is obtained by
$\rho\leftrightarrow\sigma$.
The roots belonging to each two-dimensional plane constitute the
positive roots of a two-dimensional sub-root system
of the original set of roots $\Delta$.
The only possible two-dimensional root systems  are
$A_{1}\times A_{1}$,
$A_{2}$, $B_{2}$, $G_{2}$, and $I_{2}(m)$.
 Table 2 shows the
two-dimensional sub-root systems appearing in the root systems
of finite reflection groups.
The $A_{1}\times A_{1}$ root system has been omitted since its
corresponding quantity (\ref{innerform}) is always
zero. It should be stressed that the
quantities $f_{\Theta}$ are determined by the two-dimensional sub-root
systems only and not by where they are embedded in the entire root system.

\setcounter{table}{1}
\begin{table}
\centering
   \begin{tabular}{|c|c|} \hline \label{SubrootTbl}
   Root System & Sub-root Systems \\ \hline
   $A_{r},\,r>1$ & $A_{2}$ \\ \hline
   $B_{r},\,r\ge2$ & $A_{2}$,\,$B_{2}$ \\ \hline
   $C_{r},\,r\ge2$ & $A_{2}$,\,$B_{2}$ \\ \hline
   $D_{r},\,r>3$ & $A_{2}$ \\ \hline
   $BC_{r},\,r\ge2$ & $A_{2}$,\,$B_{2}$ \\ \hline
   $E_{6}$,$E_{7}$,\,$E_{8}$ & $A_{2}$ \\ \hline
   $F_{4}$ & $A_{2}$,\,$B_{2}$ \\ \hline
   $G_{2}$ & $A_{2}$,\,$G_{2}$ \\ \hline
   $I_{2}(m)$ & $I_{2}(k)^\dagger$\\ \hline
   $H_{3}$ & $A_{2}$,\,$I_{2}(5)$ \\ \hline
   $H_{4}$ & $A_{2}$,\,$I_{2}(5)$ \\ \hline
   \end{tabular}
\caption{Two-dimensional sub-root systems.
$A_{1}\times A_{1}$ is not included.
$\dagger$: $k$ divides  $m$.}
\end{table}

Let us evaluate them in order.
We first consider the $A_2$ sub-root system with $\alpha$ and $\beta$ simple
roots, which are of the same length.
Thus the coupling dependence factorises and we obtain
\begin{eqnarray}
   2f_{A_{2}}(q)/(g^2|\rho|^2)=&&x((\alpha+\beta)\cdot q)
\,x(\alpha\cdot q)
   +x(\beta\cdot q)\,x((\alpha+\beta)\cdot q) \nonumber\\
   &&\quad -x(\alpha\cdot q)\,x(\beta\cdot q).
   \label{A2sum}
\end{eqnarray}
 For the rational potential we have, immediately
\begin{equation}
     2f_{A_{2}}(q)/(g^2|\rho|^2)={1\over{(\alpha\cdot q)
   (\beta\cdot q)((\alpha+\beta)\cdot q)}}
   \left(\beta\cdot q +\alpha\cdot q-(\alpha+\beta)\cdot q\right)=0.
\end{equation}
It is also elementary to evaluate (\ref{A2sum}) for the hyperbolic
and trigonometric potentials
by using the addition theorems for $\cot$ and $\coth$
functions. Combining the results:
\begin{equation}
   2f_{A_{2}}(q)=a^2g^2|\rho|^2\times\left\{
   \begin{array}{rl}
      0& \mbox{rational}\\
      1&\mbox{hyperbolic}\\
      -1&\mbox{trigonometric}
   \end{array}\right.
   =a^2g^2\sum_{\rho\neq\sigma\in A_{2+}}(\rho\cdot\sigma)
   \times\left\{
   \begin{array}{rl}
      0& \mbox{rational}\\
      1&\mbox{hyperbolic}\\
      -1&\mbox{trigonometric}
   \end{array}\right.
   \label{fA2val}
\end{equation}
in which $a$ is the parameter in the potential (\ref{potfun}).
The sums for the $B_2$ and $G_2$ sub-root systems may be written in terms of
the
short and long simple roots, $\alpha$ and $\beta$, respectively:
\begin{eqnarray}
   2f_{B_{2}}(q)/(g_Sg_L|\rho_L|^2) =&-& x(\alpha\cdot q)\,x
   (\beta\cdot q)
   +x((\alpha+\beta)\cdot q)\,x((2\alpha+\beta)\cdot
   q)  \nonumber\\
   &+&x(\alpha\cdot q)\,x((2\alpha+\beta)\cdot q)
   + x((\alpha+\beta)\cdot q)\, x(\beta\cdot q),
   \label{fBfunc}
\end{eqnarray}
\begin{eqnarray}
   2f_{G_{2}}(q)/(g_Sg_L|\rho_L|^2)=&-&x(\alpha\cdot
   q)\,x
   (\beta\cdot q)
   +x(\alpha\cdot
   q)\,x((3\alpha+\beta)\cdot
   q) \nonumber\\
   &+&x((2\alpha+\beta)\cdot
   q)\,x((3\alpha+\beta)\cdot
   q) + x((2\alpha+\beta)\cdot
   q)\,x((3\alpha+2\beta)\cdot
   q) \nonumber\\
   &+&x((\alpha+\beta)\cdot
   q)\,x((3\alpha+2\beta)\cdot
   q)+x((\alpha+\beta)\cdot
   q)\,
   x(\beta\cdot q).
\end{eqnarray}
The \(G_2\) root system consists of six long roots and six short
roots, and the sets of long and short roots have the same structure as the
\(A_2\) roots, scaled
and rotated by
\({\pi/6}\). The contributions from the long (short) roots only
are accounted for by \(f_{A_2}\). The above \(f_{G_2}\) denotes the
contribution from the cross terms between the long and short roots.

Again it is not difficult to evaluate
\begin{equation}
   2f_{B_{2}}(q)=a^2g_Sg_L|\rho_L|^2\times
   \left\{
   \begin{array}{rl}
      0& \mbox{rational}\\
      2&\mbox{hyperbolic}\\
      -2&\mbox{trigonometric}
   \end{array}\right.\!\!\! =
   a^2g_Sg_L\sum_{\rho\neq \sigma\in B_{2+}
   }(\rho\cdot\sigma)
   \times\left\{
   \begin{array}{rl}
      0& \mbox{rational}\\
      1&\mbox{hyperbolic}\\
      -1&\mbox{trigonometric}
   \end{array}\right.
   \label{fB2val}
\end{equation}
and
\begin{eqnarray}
   2f_{G_{2}}(q)&=&a^2g_Sg_L|\rho_L|^2\times\left\{
   \begin{array}{rl}
      0& \mbox{rational}\\
      4&\mbox{hyperbolic}\\
      -4&\mbox{trigonometric}
   \end{array}\right. \nonumber\\
   &=&
   a^2g_Sg_L\sum_{\rho,\, \sigma\in G_{2+}
 \atop{\rho:\, \mbox{\scriptsize Long},\ \sigma :\, \mbox{\scriptsize
Short} }}(\rho\cdot\sigma)
   \times\left\{
   \begin{array}{rl}
      0& \mbox{rational}\\
      1&\mbox{hyperbolic}\\
      -1&\mbox{trigonometric.}
   \end{array}\right.
   \label{fG2val}
\end{eqnarray}
The corresponding sums for the dihedral root systems
$I_2(m)$ (with rational potential) are different for
odd $m$ (simply-laced) and even $m$ (non-simply laced):
\begin{equation}
f_{I_2(m)}(q)=g^2\sum_{j\neq k}^m{(\rho_j\cdot\rho_k)
\over{(\rho_j\cdot q)(\rho_k\cdot q)}},\quad
m:\mbox{odd},
\end{equation}
\begin{eqnarray}
   f_{I_2(m)}(q)&=&g_e^2\sum_{j\neq
   k\atop{\mbox{\scriptsize even}}}^m{(\rho_j\cdot\rho_k)
   \over{(\rho_j\cdot q)(\rho_k\cdot q)}}
   +g_o^2\sum_{j\neq
   k\atop{\mbox{\scriptsize odd}}}^m{(\rho_j\cdot\rho_k)
   \over{(\rho_j\cdot q)(\rho_k\cdot q)}}\nonumber\\
   &&\quad +2g_eg_o\sum_{j:\mbox{\scriptsize even}
   \atop{k:\mbox{\scriptsize odd}}}^m{(\rho_j\cdot\rho_k)
   \over{(\rho_j\cdot q)(\rho_k\cdot q)}},\quad
   m:\mbox{even},
\end{eqnarray}
in which $g_e$ and $g_o$ are the coupling constants for the
even and odd roots. In all cases  all the roots are chosen to
have the same length $|\rho_{j}|^{2}=1$, and are
parametrised as
\begin{equation}
   \label{DihedralBasis}
   \rho_{j}=\left(\cos(j\pi/m),\sin(j\pi/m)\right),
\quad j=1,\ldots,2m.
\end{equation}
It is elementary to show that the sums vanish.
For example, for odd $m$, we have
\begin{equation}
   f_{I_2(m)}(q)={g^2\over{|q^2|}}\sum_{j\neq
   k}^m{\cos({j-k\over m}\pi)
   \over{\cos(t-{j\over m}\pi)\cos(t-{k\over m}\pi)}},\quad
   q=|q|(\cos t,\sin t),
   \label{f2imvan}
\end{equation}
which is meromorphic and periodic in $t$, with period $\pi$
and it is exponentially decreasing at $t\to\pm i\infty$.
It has possible simple poles at $t={j\pi/ m}+{\pi/2}$,
$j=1,\ldots,m$.
However, its residue at $t={j\pi/ m}+{\pi/2}$ vanishes
\[
   -\sum_{k=1}^m\!{}^\prime{\cos({j-k\over
   m}\pi)\over{\cos({\pi\over2}+{j-k\over m}\pi)}}=
   \sum_{k=1}^m\!{}^\prime\cot({j-k\over
   m}\pi)=0,
\]
in which $\sum\!{}^\prime$ means that $k=j$ term should be omitted.
Thus we find $f_{I_2(m)}(q)=0$ for odd $m$. A similar
calculation and result holds for even $m$.

The ground state energy ${\cal E}_0$ in (\ref{HBform}) depends on the
root system $\Delta$ and the choice of the potential $V$.
It has  two terms
\begin{equation}
   {\cal E}_0={\cal E}_1+{\cal E}_2.
   \label{e0e1e2}
\end{equation}
The former, ${\cal E}_1$, comes from the diagonal part coming from
the difference of $x^2$ in (\ref{HMmain}) and $V$ (see (\ref{Vxrel})) and
the additional term in (\ref{Hqcy1}) and (\ref{addharm}):
\begin{equation}
   {\cal E}_1=
   \left\{
   \begin{array}{cll}
      0&&\mbox{rational}\\
      \omega\left({r\over2}+\sum_{\rho\in\Delta_+}g_{|\rho|}\right)&&
      \mbox{rational with harmonic potential}\\
      {a^2\over2}\sum_{\rho\in\Delta_+}g_{|\rho|}^2|\rho|^2\times
      &
      \left\{
      \begin{array}{r}
          -1\\
          1
       \end{array}
       \right.
       &
       \begin{array}{l}
          \mbox{hyperbolic}\\
          \mbox{trigonometric.}
      \end{array}
   \end{array}
   \right.
\end{equation}
The latter, ${\cal E}_2$, is
the constant term coming from  (\ref{sums}). From (\ref{fA2val}),
(\ref{fB2val}) and (\ref{fG2val})
we obtain a universal formula
\begin{equation}
   {\cal E}_2={a^2\over2}\sum_{\rho\neq\sigma\in\Delta_+}
   g_{|\rho|}g_{|\sigma|}(\rho\cdot\sigma)\times
   \left\{
   \begin{array}{rl}
      0& \mbox{rational with/without harmonic potential}\\
      -1&\mbox{hyperbolic}\\
      1&\mbox{trigonometric.}
   \end{array}\right.
   \label{e2form}
\end{equation}
For actual evaluation of ${\cal E}_2$ we need to know
how many two-dimensional root systems are contained in the root system
$\Delta$.
The list is as follows:
\begin{eqnarray}
A_{r} &:& \left(\begin{array}{c}r+1 \\ 3\end{array}\right)\times A_{2},
\nonumber\\ \nonumber
B_{r} &:& 4\left(\begin{array}{c}r \\ 3\end{array}\right)\times
A_{2}^{long}
+\left(\begin{array}{c} r\\ 2\end{array}\right)\times B_{2}, \\
\nonumber
C_{r} &:& 4\left(\begin{array}{c}r \\ 3\end{array}\right)\times
A_{2}^{short}
+\left(\begin{array}{c} r\\ 2\end{array}\right)\times B_{2}, \\
\nonumber
D_{r} &:& 4\left(\begin{array}{c}r \\ 3\end{array}\right)\times A_{2},
\\
E_{6} &=& 120 \times A_{2}, \nonumber\\
E_{7} &=& 336 \times A_{2}, \nonumber\\
E_{8} &=& 1120 \times A_{2}, \nonumber\\
F_{4} &:& 16 \times A_{2}^{short} + 16 \times A_{2}^{long}
+ 18 \times B_{2}, \\ \nonumber
G_{2} &:& 1 \times G_{2},\\
BC_{r} &:& 4\left(\begin{array}{c} r \\ 3 \end{array}\right) \times
A_{2} +  \left(\begin{array}{c} r \\ 2 \end{array}\right) \times
B_{2}^{short-medium}
+ \left(\begin{array}{c} r \\ 2 \end{array}\right) \times
B_{2}^{medium-long}.\nonumber
\end{eqnarray}
The non-crystallographic root systems are not listed since the constant
terms are zero in these cases.
We list ${\cal E}_2(\Delta, \mbox{trig.})/a^2$ for various root systems
in Table 3.
\begin{equation}
\begin{tabular}{|c|c|}
\hline
$\Delta$ & ${\cal E}_2(\Delta, \mbox{trig.})/a^2$ \\ \hline
$A_{r}$ & ${|\rho|^{2} g^{2}\over 2}\left(
\begin{array}{c} r+1\\ 3 \end{array}\right)$ \\ \hline
$B_{r}$ & $2|\rho_{S}|^{2}g_{L}\left[
2\left(\begin{array}{c} r \\ 3 \end{array}\right) g_{L}
+\left(\begin{array}{c} r \\ 2 \end{array}\right)g_{S}\right]$
\\ \hline
$C_{r}$ & $2|\rho_{S}|^{2}g_{S}\left[
\left(\begin{array}{c} r\\ 3 \end{array}\right) g_{S}
+\left(\begin{array}{c} r \\ 2 \end{array}\right) g_{L}\right]$ \\
\hline
$D_{r}$ & $2|\rho|^{2}\left(\begin{array}{c} r \\ 3
\end{array}\right)
g^{2}$ \\ \hline
$E_{6}$ & $60 |\rho|^{2} g^{2}$ \\ \hline
$E_{7}$ & ${118} |\rho|^{2} g^{2}$ \\ \hline
$E_{8}$ & $560 |\rho|^{2} g^{2}$ \\ \hline
$F_{4}$ & $4|\rho_{S}|^{2}\left[2g_{S}^{2}
+4 g_{L}^{2} + 9 g_{S} g_{L}\right]$ \\ \hline
$G_{2}$ & ${|\rho_{S}|^{2}}\left[g_{S}^{2}
+3g_{L}^{2}+ 12g_{S}g_{L}\right]/2$ \\ \hline
$BC_r$ & $2|\rho_{S}|^{2}
g_{M}\left[2\left(\begin{array}{c} r \\ 3 \end{array}\right) g_{M}
+ \left(\begin{array}{c} r \\ 2 \end{array}\right) g_{S}
+2 \left(\begin{array}{c} r \\ 2 \end{array}\right) g_{L} \right]$
\\ \hline
\end{tabular}
\label{e2list}
\end{equation}
\begin{center}
Table 3:\ The part of the ground state energy ${\cal E}_2 /a^2$ for the
trigonometric potential.
\end{center}

\bigskip
We arrive at the following explicit forms of the bosonic
and fermionic Hamiltonians
${\cal H}_B$ (\ref{HBform}) and ${\cal H}_F$ (\ref{HSCF}):
\begin{equation}
   {\cal H}_B = {1\over 2} p^{2} + {1\over2}\sum_{\rho\in\Delta_+}
   g_{|\rho|}(g_{|\rho|}-1) |\rho|^{2}
   \,V(\rho\cdot q)+({\omega^2\over2}q^2)-{\cal E}_0,
   \label{HBexp}
\end{equation}
\begin{equation}
   {\cal H}_F=\sum_{\rho\in\Delta_+}
   g_{|\rho|}(\rho\cdot\psi^*)(\rho\cdot\psi)V(\rho\cdot q)+
   (\omega\psi^*\cdot\psi).
   \label{HFexp}
\end{equation}
For the hyperbolic and trigonometric cases ${\cal E}_0 = {\cal E}_1+{\cal
E}_2$
is expressed succinctly as:
\begin{equation}
   {\cal E}_0=
   {a^2\over2}\left(\sum_{\rho\in\Delta_+}
   g_{|\rho|}\rho\right)^2
   \times
   \left\{
   \begin{array}{cl}
      -1&\mbox{hyperbolic}\\
      1&\mbox{trigonometric.}
   \end{array}
   \right.
   \label{e0fom}
\end{equation}
For the rational potential cases the ground state energy ${\cal E}_0$ is
\begin{equation}
{\cal E}_0=
\left\{
\begin{array}{cl}
0&\mbox{rational}\\
\omega\left({r\over2}+\sum_{\rho\in\Delta_+}g_{|\rho|}\right)&\mbox{rational
with harmonic potential.}
\end{array}
\right.
\label{ratgren}
\end{equation}

The bosonic Hamiltonian ${\cal H}_B$ (\ref{HBexp}) has the same form as
the classical Hamiltonian (\ref{cCMHamiltonian}) with only one replacement
\begin{equation}
   g_{|\rho|}^2\to g_{|\rho|}(g_{|\rho|}-1),
\end{equation}
which is essential for quantum integrability as we will see shortly.
It should be remarked that the mechanism which guarantees
\begin{equation}
   {\cal H}_{C}={1\over 2}\sum_{j=1}^{r}\left(p_{j}^{2}+\left({\partial W
   \over{\partial q_{j}}}\right)^{2}\right)+const.
\end{equation}
is the same one that guarantees the consistency of the classical Lax
equation
(\ref{LaxEquation}) \cite{bcs2}.
The same mechanism plays an important role in the consistency of the
quantum conserved quantities, as we shall see in section \ref{qlax}.

\subsection{Vacuum or ground state}
Supersymmetric quantum mechanics provides the easiest way to construct the
supersymmetric vacuum, which also gives the ground state energy and
eigenfunction of
the pure bosonic theory.
The supersymmetric vacuum state $|vac\rangle$ is annihilated
by the supercharges
\begin{equation}
   {\cal Q}|vac\rangle={\cal Q}^*|vac\rangle=0,
   \end{equation}
therefore it is an eigenstate of the supersymmetric
Hamiltonian with zero energy
   \begin{equation}
   {\cal H}_{SUSY}|vac\rangle=0.
   \label{HSCgreq}
\end{equation}
In order to express $|vac\rangle$ explicitly, let us introduce the state
$|0\rangle$ which is annihilated by all of the fermionic
annihilation operators:
\begin{equation}
   \psi_j|0\rangle=0,\quad j=1,\ldots,r.
\end{equation}
Let us suppose that
\begin{equation}
   |vac\rangle=\Phi_0(q)|0\rangle,
\end{equation}
in which $\Phi_0(q)$ is yet to be determined.
Then it satisfies
\(
{\cal Q}^*|vac\rangle=0
\)
trivially. The other condition
\(
{\cal Q}|vac\rangle=0
\)
is fulfilled if \(\Phi_0\) satisfies
\begin{equation}
   \left(p_j+i{\partial W\over{\partial
   q_j}}\right)\Phi_0=0,\quad  j=1,\ldots,r.
   \label{Weq}
\end{equation}
A solution of (\ref{Weq}) is given simply by
\begin{equation}
   \Phi_0(q)=e^{W(q)}=\prod_{\rho\in\Delta_+}
   |w(\rho\cdot q)|^{g_{|\rho|}}\,e^{-{\omega\over2}q^2},
   \label{grsol}
\end{equation}
which is real and Coxeter invariant (\ref{Coxinv}).
The exponential factor $e^{-{\omega\over2}q^2}$
exists only for the rational
potential case with the harmonic confining force.

By substituting the above solution (\ref{grsol}) into (\ref{HSCgreq})
and using the decomposition of the supersymmetric Hamiltonian (\ref{HBHF})
together with
\[
   {\cal H}_F|vac\rangle=0,
\]
we obtain from ${\cal H}_B|vac\rangle=0$
\begin{equation}
   \left({1\over 2} p^{2} + {1\over2}\sum_{\rho\in\Delta_+}
   g_{|\rho|}(g_{|\rho|}-1) |\rho|^{2}
   \,V(\rho\cdot q)+({\omega^2\over2}q^2)\right)\,e^W={\cal E}_0\,e^W.
   \label{purbossol}
\end{equation}
In other words, the above solution (\ref{grsol}) provides a ground state
with energy ${\cal E}_0$ of the pure bosonic model with Hamiltonian
${\cal H}_C+{\cal H}_{qc}$. It should be stressed that  ${\cal E}_0$ is
determined purely algebraically using (\ref{e0e1e2})--(\ref{e2list}),
without really applying the operator on the left hand side of
(\ref{purbossol}) to the solution.
In fact, one would need essentially the same calculation as above
to show that \(e^{W}\) is an eigenstate by direct application of the
Hamiltonian operator. Supersymmetry provides the simplest means to
assert that it is the ground state.
This type of ground state has been known for some time.
It is derived by various methods, see for example \cite{CalMo,OP2}, and
also by using supersymmetric quantum mechanics
for the models based on classical root systems \cite{FM,ShaSu}.
Needless to say, our solution (\ref{grsol}) provides a {\em universal}
ground state solution for all the models considered in this paper.

The other states of the bosonic models can be obtained as
eigenfunctions of a differential operator $\widetilde{\cal H}_B$ obtained
from
${\cal
H}_B$ by a similarity transformation:
\begin{eqnarray}
   \widetilde{\cal H}_B&=&
   e^{-W}\,{\cal H}_B\,e^W\nonumber\\
   &=&e^{-W}\left({1\over 2} p^{2} +
   {1\over2}\sum_{\rho\in\Delta_+}
   g_{|\rho|}(g_{|\rho|}-1) |\rho|^{2}
   \,V(\rho\cdot q)+({\omega^2\over2}q^2)-{\cal E}_0\right)\,e^W,
\end{eqnarray}
\begin{equation}
   \widetilde{\cal H}_B\phi_{\lambda}=\lambda\phi_{\lambda} \quad
   \Longleftrightarrow
   {\cal H}_B\,\phi_{\lambda}\,e^W=\lambda\phi_{\lambda}\,e^W.
   \label{neweq}
\end{equation}
Obviously we have
\begin{equation}
   \int_{PW\,(PW_T)}e^{2W(q)}\,dq=\left\{
   \begin{array}{cl}
      \infty&\mbox{: rational and hyperbolic}\\
      \mbox{finite}&\mbox{: trigonometric and rational with the harmonic
   potential},
   \end{array}
   \right.
\end{equation}
in which $PW$ and $PW_T$ denote that the integration is over the regions
defined in (\ref{PW}) and (\ref{PWT}).
It should be remarked that the `ground state' wavefunctions and
`ground state' energies in the non-normalisable cases ({\em i.e.} the
rational and hyperbolic potentials and, in particular, the negative
`ground state' energy of the latter)
should not be taken at face value.
In the rational (hyperbolic) case the wavefunction
$\Phi_0(q)=e^{W(q)}$ diverges
polynomially (exponentially) for $\alpha_h\cdot q\to+\infty$.
A similar and better-known situation arises in the quantum mechanics of
a free particle in one-dimension: ${\cal H}=p^2/2$.
It has an exponential `eigenstate' with a negative energy:
\[
 {\cal H}\phi_0(q)=-{k^2\over2}\phi_0(q),\quad
 \phi_0(q)=e^{kq},\quad
 k\in{\bf R}.
\]

Naturally,  most existing results in quantum Calogero-Moser models are
for the models with normalisable states. There are also some results
for the rational and hyperbolic models \cite{CalMo,Dunk,AhNa}.
We will not discuss the eigenstates and spectra of (\ref{neweq}) further.
In the rest of this paper we will concentrate on the
integrability structure of the quantum pure bosonic system (\ref{HBexp}).

\section{Quantum Lax Pair Operators}
\label{qlax}
\setcounter{equation}{0}
In this section we present the formulation of the quantum Lax
pair operators, which enables us to construct the quantum conserved
quantities  for the Calogero-Moser models based on all of the root systems
(crystallographic and non-crystallographic) and for all of the degenerate
potentials, as in the classical case given in section 2.
We believe such a universal construction of the quantum Calogero-Moser Lax
pair is new.
We will write down the quantum equations of motion of the
Calogero-Moser models in an equivalent matrix form, whose matrix elements
are
quantum operators. Surprisingly the difference between the classical and
quantum Lax pair operators is very small, as we will see below.

The Lax pair operators for {\em classical} Calogero-Moser models
imply the quantum equations of motion, if they are interpreted as
quantum
operators.
We start from the classical Hamiltonian

\begin{equation}
   \label{cCMHamil2}
   \mathcal{H}_C = {1\over 2} p^{2} + {1\over2}\sum_{\rho\in\Delta_+}
   {g_{|\rho|}^{2} |\rho|^{2}}
   \,V(\rho\cdot q),
\end{equation}
and consider its action as the quantum evolution operator.
This is not at all strange, since the solvable Hamiltonians of the harmonic
oscillator and the hydrogen atom are the same at the classical and quantum
levels. The canonical equations of motion and quantum Heisenberg equations
of
motion are formally identical:
\begin{eqnarray}
   \dot{q}&=&\{q,{\cal H}_C\}=i[{\cal H}_C,q]=p,
   \label{caneq1}\\
   \dot{p}&=&\{p,{\cal H}_C\}=i[{\cal
   H}_C,p]=-{1\over2}\sum_{\rho\in\Delta_+}
   {g_{|\rho|}^{2}|\rho|^{2}}
   \,V^\prime(\rho\cdot q)\,\rho.
   \label{caneq2}
\end{eqnarray}
As shown in section \ref{cal-mo},  the {\em classical}
equations of motion are equivalent to the
Lax form (\ref{LaxEquation})
\begin{equation}
   \label{LaxEquation2}
     {d\over dt}{L}=[L,\widetilde{M}],
\end{equation}
which is divided into two parts as
\begin{eqnarray}
   {d\over dt}X &=& [p\cdot\hat{H},\widetilde{M}],
   \label{lax1}\\
   {d\over dt}(p\cdot\hat{H}) &=&
   [X,\widetilde{M}].
   \label{lax2}
\end{eqnarray}

The second equation (\ref{lax2}) corresponds to (\ref{caneq2}).
Since only the $q$ operators appear on the right hand side of
(\ref{lax2}), the quantum commutator $[X,\widetilde{M}]$ is the same as
the classical one, depending only on the matrix structure.
We still have to consider the first equation (\ref{lax1}) which
could be different from the classical one when $p$ and $q$ are
non-commuting.
For this purpose, let us evaluate ${dx(\rho\cdot q)/dt}$ quantum
mechanically:
\begin{eqnarray}
   &&{d\over{dt}}x(\rho\cdot q)\nonumber\\
   &=&i\left[{\cal H}_C, x(\rho\cdot q)\right]
   =i\left[{p^2/2},x(\rho\cdot q)\right]\nonumber\\
   &=&{1\over2}\left(
    \phantom{\mbox{\Large H}}\hspace{-10pt}
   (p\cdot\rho)\, y(\rho\cdot q)
   +y(\rho\cdot q)\,(p\cdot\rho)\right),
   \label{qcanq}
\end{eqnarray}
in which $x^\prime=y$ is used. The right hand side is Weyl
(symmetrically) ordered.

Next let us evaluate the matrix element
\[
   [p\cdot\hat{H}, \widetilde{M}]_{\mu\nu}=\left[p\cdot\hat{H},
   \,{(i/2)}\sum_{\rho\in\Delta_{+}}g_{|\rho|}
   \,{|\rho|^2}
     \,y (\rho\cdot q)\,\hat{s}_{\rho}\right]_{\mu\nu},\quad
\mu,\nu\in{\cal R}
\]
quantum mechanically. We find
\begin{eqnarray}
   &&[p\cdot\hat{H},\widetilde{M}]_{\mu\nu}\nonumber\\
   &=&\left((p\cdot\hat{H}){(i/2)}\sum_{\rho\in\Delta_{+}}g_{|\rho|}
   |\rho|^2\,y
      (\rho\cdot q)\,
      \hat{s}_{\rho}- {(i/2)}\sum_{\rho\in\Delta_{+}}g_{|\rho|}|\rho|^2
   \,y(\rho\cdot q)\,\hat{s}_{\rho}
     \, (p\cdot\hat{H})\right)_{\mu\nu}\nonumber\\
   &=&{(i/2)}\sum_{\rho\in\Delta_{+}}g_{|\rho|}|\rho|^2\left(
   \phantom{\mbox{\huge H}}\hspace{-15pt}
   p\cdot s_{\rho}(\nu)\,
   \,y(\rho\cdot q)
   -\,y(\rho\cdot q)
      p\cdot\nu\right)(\hat{s}_{\rho})_{\mu\nu}.
      \label{xmmid}
\end{eqnarray}
Note that
\[
   s_{\rho}(\nu)=\nu-(\rho^\vee\!\!\cdot\nu)\rho,\quad
   p\cdot s_{\rho}(\nu)=p\cdot\nu-(\rho^\vee\!\!\cdot\nu)p\cdot\rho,
\]
and it is easy to see that
\begin{eqnarray}
   &&p\cdot\nu {(i/2)}\sum_{\rho\in\Delta_{+}}g_{|\rho|}|\rho|^2
   \,y(\rho\cdot q)\,
    -{(i/2)}\sum_{\rho\in\Delta_{+}}g_{|\rho|}|\rho|^2
   \,y(\rho\cdot q)\,
      p\cdot\nu \nonumber\\
   &&\quad= {(1/2)}\sum_{\rho\in\Delta_{+}}g_{|\rho|}|\rho|^2
   \,y^\prime(\rho\cdot q)\,
      \rho\cdot\nu.
\end{eqnarray}
Thus we arrive at
\begin{eqnarray}
   &&[p\cdot\hat{H},\widetilde{M}]_{\mu\nu}\nonumber\\
   &=&(1/2)\sum_{\rho\in\Delta_{+}}g_{|\rho|}
   |\rho|^2\left(\phantom{\mbox{\huge H}}\hspace{-15pt}
   -i(\rho^\vee\!\!\cdot\nu)
   (p\cdot\rho)\,y(\rho\cdot q)\,
     +
   \,y^\prime(\rho\cdot q)\,
      \rho\cdot\nu\right)(\hat{s}_{\rho})_{\mu\nu}.
\end{eqnarray}
At this point we split
$p\cdot\rho={(1/2)}p\cdot\rho+{(1/2)}p\cdot\rho$
and apply the second momentum operator to the function $y$
\begin{equation}
(p\cdot\rho)\, y(\rho\cdot q)\nonumber\\
   =-i{|\rho|^2}\,y^\prime
   (\rho\cdot q)
   + \,y(\rho\cdot q)
   \,(p\cdot\rho).
\end{equation}
Thus we arrive at
\begin{eqnarray}
   &&[p\cdot\hat{H},\widetilde{M}]_{\mu\nu}\nonumber\\
   &=&\left(-{i/4}\sum_{\rho\in\Delta_{+}}g_{|\rho|}
   |\rho|^2(\rho^\vee\!\!\cdot\nu)
   \left(\phantom{\mbox{\huge H}}\hspace{-15pt}
(p\cdot\rho)\,y(\rho\cdot q)+y(\rho\cdot q)
      (p\cdot\rho)\right)\right.
       \nonumber\\
   &&\left. + \sum_{\rho\in\Delta_{+}}g_{|\rho|}
      \left(\phantom{\mbox{\Large H}}\hspace{-13pt}
      -\rho^\vee\!\!\cdot\nu\,|\rho|^2/4+\rho\cdot\nu/2\right)
      y^\prime(\rho\cdot q)\right)
      (\hat{s}_{\rho})_{\mu\nu}.
\end{eqnarray}
The second line vanishes, since $\rho^\vee=2\rho/|\rho|^2$.
By using the formulas
\[
   \rho^\vee\!\!\cdot s_{\rho}(\nu)=\rho^\vee\!\!\cdot
   (\nu-(\rho^{\vee}\!\!\cdot\nu)\rho)=-\rho^\vee\!\!\cdot\nu, \quad
   \rho^\vee\!\!\cdot\rho=2,
\]
we obtain
\begin{eqnarray}
   &&[p\cdot\hat{H},\widetilde{M}]_{\mu\nu}\nonumber\\
   &=&{i/2}\sum_{\rho\in\Delta_{+}}g_{|\rho|}(\rho\cdot
   s_{\rho}(\nu))
   \left(\phantom{\mbox{\huge H}}\hspace{-15pt}
      (p\cdot\rho)\,y(\rho\cdot q)+ y(\rho\cdot q)
      (p\cdot\rho)\right)
      (\hat{s}_{\rho})_{\mu\nu} \nonumber\\
   &=&{i/2}\sum_{\rho\in\Delta_{+}}g_{|\rho|}(\rho\cdot
   \hat{H})
   \left(\phantom{\mbox{\huge H}}\hspace{-15pt}
      (p\cdot\rho)\,y(\rho\cdot q) + y(\rho\cdot q)
      (p\cdot\rho)\right)
      (\hat{s}_{\rho})_{\mu\nu} \nonumber\\
   &=&\left(i\sum_{\rho\in\Delta_{+}}g_{|\rho|}
   (\rho\cdot\hat{H}){d\over{dt}}x(\rho\cdot q)
      \hat{s}_{\rho}\right)_{\mu\nu} \nonumber\\
   &=&{d\over{dt}}X_{\mu\nu}=i[{\cal H}_C,X_{\mu\nu}].
\end{eqnarray}

Thus we have established that the first Lax equation has the same form at
the quantum and classical levels:
\begin{equation}
   {d\over{dt}}X=i[{\cal H}_C, X]=[p\cdot\hat{H},\widetilde{M}]
   \end{equation}
   and that the quantum Lax equation as a whole has the same form as the
classical one
   \begin{equation}
   {d\over{dt}}L=i[{\cal H}_C, L]=[L,\widetilde{M}]
\end{equation}
or to be more precise
\begin{equation}
   i[{\cal H}_C,
   L_{\mu\nu}]=\sum_{\lambda\in{\cal
   R}}\left(L_{\mu\lambda}\widetilde{M}_{\lambda\nu}-
   \widetilde{M}_{\mu\lambda}L_{\lambda\nu}\right),\qquad
   \mu,\nu\in{\cal R}.
\end{equation}
Similarly we obtain for the rational model with the harmonic force
\begin{equation}
   i[{\cal H}_{C\omega},
   L^\pm_{\mu\nu}]=\sum_{\lambda\in{\cal
   R}}\left(L^\pm_{\mu\lambda}\widetilde{M}_{\lambda\nu}-
   \widetilde{M}_{\mu\lambda}L^\pm_{\lambda\nu}\right)\pm i\omega
   L^\pm_{\mu\nu},\qquad
   \mu,\nu\in{\cal R}.
\end{equation}
\noindent From these it is straightforward to derive (recalling the
definitions (\ref{defcalL}))
\begin{eqnarray}
   {d\over{dt}}L^n&=&i[{\cal H}_C,L^n]=[L^n,\widetilde{M}],\quad
   n=1,2,\ldots ,\\
   {d\over{dt}}{\cal L}_k^n&=&i[{\cal
   H}_{C\omega},{\cal L}_k^n]=[{\cal L}_k^n,\widetilde{M}],\quad
   k=1,2,\quad n=1,2,\ldots .
\end{eqnarray}
However, the parallelism between the classical and quantum Lax equations
ends here. These equations do not imply that Tr$L^n$ and Tr${\cal L}_k^n$
are
conserved.
This is because the matrix elements of the quantum $L$ and $\widetilde{M}$
operators do
not commute and the cyclicity of the matrix trace
is broken.

\bigskip
The remedy is simple.
We adopt the bosonic Hamiltonian ${\cal H}_B$ (\ref{HBexp}) obtained from
the superpotential in the previous section and the Lax pair
$L$ and $M$ instead of $L$ and $\widetilde{M}$.
The quantum Lax bracket
\begin{equation}
   [L,M]=[L,\widetilde{M}+i{\cal D}\times I]
   =[L,\widetilde{M}]+[p\cdot\hat{H},i{\cal D}\times I]
\end{equation}
is {\em different} from the classical one, since the last term is
no longer vanishing:
\begin{equation}
   [p\cdot\hat{H},i{\cal D}\times I]={1\over2}\sum_{\rho\in\Delta_+}
   {g_{|\rho|}|\rho|^{2}}
   \,V^\prime(\rho\cdot q)\,\rho\cdot\hat{H}.
\end{equation}
This term provides the necessary difference between the equations of motion
of ${\cal H}_C$ and ${\cal H}_B$. Thus we have established
for the Hamiltonian ${\cal H}_B$
\begin{eqnarray}
   {d\over{dt}}(L^n)_{\mu\nu}&=&i[{\cal H}_B,(L^n)_{\mu\nu}]
   =[L^n,M]_{\mu\nu}
   \nonumber\\
   &=&\sum_{\lambda\in{\cal R}}\left(\phantom{\mbox{\huge
   H}}\hspace{-15pt}
   (L^n)_{\mu\lambda}M_{\lambda\nu}-
   M_{\mu\lambda}(L^n)_{\lambda\nu}\right),\quad
   n=1,\ldots ,
\label{qlaxL}\\
   {d\over{dt}}({\cal L}_k^n)_{\mu\nu}&=&i\left[{\cal
   H}_{B\omega},({\cal L}_k^n)_{\mu\nu}\right]=\left[{\cal
L}_k^n,M\right]_{\mu\nu},\quad
   k=1,2,\quad n=1,2,\ldots .
\label{qlaxLcal}
\end{eqnarray}
The above equations are operator equations, therefore they are
valid for any  Calogero-Moser models with any potentials and  any
representations of the Lax
pairs.

We define {\em quantum conserved quantities} as the {\em total sum}
$(\mbox{Ts})$ of all
matrix elements of $L^n$ (${\cal L}_k^n$, $k=1,2$):
\begin{eqnarray}
   Q_n&=&\mbox{Ts}(L^n)\equiv\sum_{\mu,\nu\in{\cal R}}(L^n)_{\mu\nu},\quad
   n=1,\ldots,
\label{qconv}\\
   Q^{\omega, k}_n&=&\mbox{Ts}({\cal L}_k^n)\equiv
   \sum_{\mu,\nu\in{\cal R}}({\cal L}_k^n)_{\mu\nu},\quad
   k=1,2,\quad
   n=1,\ldots.
\label{qconvcal}
\end{eqnarray}
They are conserved thanks to the property of the $M$ operator
(\ref{sumMzero}):
\[
   \sum_{\mu\in{\cal R}}M_{\mu\nu}=
   \sum_{\nu\in{\cal R}}M_{\mu\nu}=0.
\]
Such quantum conserved quantities have been previously reported for some
models based on $A_r$ root systems \cite{ShaSu,UjWa}.
It should be remarked that {Ts}$({\cal L}_2^n)$ is no longer the same as
{Ts}$({\cal L}_1^n)$ due to quantum corrections.
As we will show at the end  this section, the quantum Hamiltonian
{Ts}$({\cal L}_2)$ differs from {Ts}$({\cal L}_1)$ by a constant.

Next we show that the  quantum Hamiltonian ${\cal H}_B$ is obtained by
taking the total sum   of $L^2$ in a representation ${\cal R}$:
\[
    {\cal H}_B\propto \mbox{Ts}(L^2).
\]
This is a necessary condition for the internal consistency of the quantum
Lax pair operator formalism.
We start from
\[
   L^2=\left(p\cdot\hat{H}\right)^2+
       \left(p\cdot\hat{H}X+X\,p\cdot\hat{H}\right)+X^2.
\]
For the diagonal operator $\hat{H}$,  Ts and Tr are the same and
the first term,
$(p\cdot\hat{H})^2$, gives as in the classical theory $p^2 C_{\cal R}$.
The next term reads
\begin{eqnarray}
   &&p\cdot\hat{H}X+X\,p\cdot\hat{H}
   \nonumber\\&=&
   p\cdot\hat{H}\left(i\sum_{\rho\in\Delta_{+}}g_{|\rho|}
      \,\,(\rho\cdot\hat{H})\,x(\rho\cdot q)\,\hat{s}_{\rho}\right)
      +\left(i\sum_{\rho\in\Delta_{+}}g_{|\rho|}
      \,\,(\rho\cdot\hat{H})\,x(\rho\cdot q)\,\hat{s}_{\rho}\right)
      p\cdot\hat{H}
      \nonumber\\
   &=&\sum_{\rho\in\Delta_+}g_{|\rho|}y(\rho\cdot q)(\rho\cdot\hat{H})^2
   \hat{s}_{\rho}
   \nonumber\\
   &&\quad +i\sum_{\rho\in\Delta_+}g_{|\rho|}x(\rho\cdot
q)(\rho\cdot\hat{H})
   \left((p\cdot\hat{H})\,\hat{s}_{\rho}+ \hat{s}_{\rho}\,(p\cdot\hat{H})
   \right).
   \label{pHS}
\end{eqnarray}
Here, the first term on the right hand side of (\ref{pHS}) gives the
same expression as ${\cal H}_{qc}$, (\ref{Hqcy2}),
as we have
\begin{equation}
   \sum_{\mu,\nu\in{\cal R}}\left((\rho\cdot\hat{H})^2\hat{s}_{\rho}
   \right)_{\mu\nu}=
   \sum_{\mu,\nu}(\rho\cdot\mu)^2(\hat{s}_{\rho})_{\mu\nu}=
   \sum_{\mu}(\rho\cdot\mu)^2=C_{\cal R}|\rho|^2,
\end{equation}
in which the formula $\sum_{\nu\in{\cal R}}(\hat{s}_{\rho})_{\mu\nu}=1$,
 (\ref{ssumform}),
is used. The second sum in (\ref{pHS}) vanishes, since
we have
\begin{eqnarray}
   && \sum_{\mu,\nu}\left(\phantom{\mbox{\huge H}}\hspace{-15pt}
   \rho\cdot\hat{H}\left(p\cdot\hat{H}\hat{s}_{\rho}
   +\hat{s}_{\rho}p\cdot\hat{H}\right)\right)_{\mu\nu}
   \nonumber\\
   &&=\sum_{\mu,\nu}\rho\cdot s_{\rho}(\nu)\left(
   \phantom{\mbox{\Large H}}\hspace{-10pt}
    p\cdot
    s_{\rho}(\nu)+p\cdot\nu\right)
   (\hat{s}_{\rho})_{\mu\nu}
   \nonumber\\
   &&=-\sum_{\nu}(\rho\cdot\nu)\,p\cdot\left(
   2\nu-(\rho^\vee\!\!\cdot\nu)\rho\right)
    =C_{\cal R}(-2\rho\cdot p+\rho^\vee\!\!\cdot\rho \rho\cdot p)
   \nonumber\\
   &&=0,
\end{eqnarray}
in which (\ref{ssumform}) is used again. Finally we show that
Ts$X^2=$Tr$X^2$, which is rather non-trivial since
the off-diagonal terms
$(X^2)_{\mu\nu}$ are generally non-vanishing:
\begin{equation}
   (X^2)_{\mu\nu}=
   -\sum_{\rho,\sigma}g_{|\rho|}g_{|\sigma|}
   x(\rho\cdot q)x(\sigma\cdot q)
   \left(\rho\cdot\hat{H}\hat{s}_{\rho}\ \sigma\cdot\hat{H}
   \hat{s}_{\sigma}\right)_{\mu\nu},
\end{equation}
 in which
\[
   \left(\rho\cdot\hat{H}\hat{s}_{\rho}\ \sigma\cdot\hat{H}
   \hat{s}_{\sigma}\right)_{\mu\nu}=\rho\cdot(s_{\sigma}(\nu))
   \ \sigma\cdot\nu\,(\hat{s}_{\rho}\hat{s}_{\sigma})_{\mu\nu}.
\]
Since $\sum_{\mu}(\hat{s}_{\rho}\hat{s}_{\sigma})_{\mu\nu}=1$
for $\rho=\sigma$ and $\rho\neq\sigma$, we obtain
\begin{equation}
   \sum_{\mu,\nu}\left(\rho\cdot\hat{H}\hat{s}_{\rho}\  \sigma\cdot\hat{H}
   \hat{s}_{\sigma}\right)_{\mu\nu}=-C_{\cal R}\rho\cdot\sigma,
\end{equation}
hence
\begin{equation}
   \sum_{\mu,\nu\in{\cal R}}(X^2)_{\mu\nu}=C_{\cal R}
   \sum_{\rho,\sigma}g_{|\rho|}g_{|\sigma|}(\rho\cdot\sigma)
   x(\rho\cdot q)x(\sigma\cdot q),
\end{equation}
which is proportional to the $(\partial W/\partial q)^2$ term in
(\ref{HMmain}).
Thus we have established the announced result for the models
without the harmonic potential:
\begin{equation}
   {1\over{2C_{\cal R}}}\mbox{Ts}(L^2)={\cal H}_B=
   {1\over 2} p^{2} + {1\over2}\sum_{\rho\in\Delta_+}
   g_{|\rho|}(g_{|\rho|}-1) |\rho|^{2}
   \,V(\rho\cdot q)-{\cal E}_0.
\end{equation}
It should be emphasised that the mechanism which ensures the
equality Ts$X^2$=Tr$X^2$ is the same one which allows the introduction
of supersymmetry in section \ref{susy}.

Finally we establish that the Hamiltonian of the rational model
with the harmonic potential is obtained in a similar way.
To do this, we show that
\begin{equation}
   {\cal H}_B \propto\sum_{\mu,\nu\in{\cal R}}(L^+L^-)_{\mu\nu}
   =\sum_{\mu,\nu\in{\cal R}}(L^-L^+)_{\mu\nu} + const,
   \label{harham}
\end{equation}
in which
\[
   L^{\pm}=L\pm i\omega Q, \quad Q=q\cdot\hat{H}.
\]
We have
\begin{eqnarray}
   L^{\pm}L^{\mp}&=&
   (L\pm i\omega Q)(L\mp i\omega Q)
   \nonumber\\
   &=&L^2+\omega^2Q^2 \pm i\omega(QL-LQ),
\label{LpmLmp}
\end{eqnarray}
and
\begin{eqnarray}
   QL-LQ&=&
   Q(p\cdot\hat{H}+X)-(p\cdot\hat{H}+X)Q
   \nonumber\\
   &=&i\hat{H}_j\hat{H}_k\delta_{jk}+QX-XQ.
\end{eqnarray}
The second term is the same as in the classical theory
\begin{equation}
   QX-XQ=i\sum_{\rho\in\Delta_+}g_{|\rho|}(\rho\cdot\hat{H})
   (\rho^\vee\!\!\cdot\hat{H})\hat{s}_{\rho}.
\end{equation}
Thus we arrive,
using (\ref{LpmLmp}), at
\begin{equation}
   {1\over{2C_{\cal R}}}\sum_{\mu,\nu\in{\cal R}}
   \left(L^{\pm}L^{\mp}\right)_{\mu\nu}=
   {1\over 2} p^{2} + {1\over2}\sum_{\rho\in\Delta_+}
   g_{|\rho|}(g_{|\rho|}-1)
   \,{|\rho|^{2}\over{(\rho\cdot q)^2}}+{\omega^2\over2}q^2
   \mp \omega({r\over2}+\sum_{\rho\in\Delta_+}g_{|\rho|}),
\end{equation}
which confirms (\ref{harham}).
The constant term is the ground state energy (\ref{ratgren}).


\section{Comments and Discussion}
\label{comdis}
\setcounter{equation}{0}
In this paper we have established in an elementary way how all of the
Calogero-Moser models with  degenerate potentials can be
supersymmetrised.
As a by-product, universal formulas for the ground state energies and
wavefunctions of the original ({\em i.e.} non-supersymmetric)
quantum Calogero-Moser models are obtained.
We have also given quantum Lax pair operators for these models and
derived quantum conserved quantities.
These results would constitute a good starting point for the systematic
study
of quantum Calogero-Moser models, in particular, those based on the
exceptional and non-crystallographic root systems.
Besides the seminal work by Dunkl on the models based on the dihedral
groups \cite{Dunk}, there are many works on the quantum $G_2$ model, the
exceptional Calogero-Moser model with the fewest ({\em i.e.} 2) degrees of
freedom
\cite{G2} and some on the $H_3$ and $F_4$ models \cite{exceps}.
One advantage of our formulation is its universality and another is that
it is independent of any specific choice of the realisation of the root
systems.

Another merit of constructing quantum conserved quantities in terms of
quantum Lax operators is that it becomes obvious that these conserved
quantities are operators acting on functions in the configuration
space, that is, either a Weyl chamber (\ref{PW}) or a Weyl alcove
(\ref{PWT}).
The quantum theory we are discussing is the so-called first quantised
theory.
That is, the notion of identical particles and the associated statistics
is non-existent.
The symmetry properties of quantum solutions with respect to the action
of the reflection operators will be discussed elsewhere.

For technical reasons, we have not developed the corresponding
theory of commuting differential operators for these models
\cite{BFV,Cher,Heck2,HeOp}.  As is well-known, the theory of commuting
differential operators would provide another method for constructing
quantum conserved quantities.
Analysis of the spectrum and
eigenfunctions of Calogero-Moser models based on commuting
differential operators
\cite{ViLa,BaFo,UjWa}, shape-invariance \cite{HeOp,Eft,GrPan,Kha} and
quantum Lax pairs will be published elsewhere.

\section*{Acknowledgements}
\setcounter{equation}{0}
We thank S.\,P.\, Khastgir for
useful discussions and S.\, Ruijsenaars for bringing \cite{HeOp} to our
attention. R.\,S. thanks M.\, Wadati, K.\, Hikami and  Y.\, Komori for
useful discussions. N.\,S.\,M. thanks the YukawaInstitute for hospitality.
This work is partially supported  by the Grant-in-aid from
the  Japanese Ministry of Education, Science and Culture,
priority area (\#707)  ``Supersymmetry and unified theory of elementary
particles". A.\,J.\,B. is supported by the Japan Society for the
Promotion of Science.  A.\,J.\,B. is also supported by the U.S. National
Science Foundation under grant no. 9703595.



\end{document}